\documentclass[aps,prl,floatfix,twocolumn,showpacs,nofitinbib,hyperref=pdftex,citeautoscript]{revtex4-1}
\usepackage{epsfig}
\usepackage{amsmath}
\usepackage{amssymb}
\usepackage{amsfonts}
\usepackage{ dsfont }
\usepackage{ comment,color }
\usepackage{lipsum}
\usepackage{hyperref,mathtools}

\hypersetup{colorlinks=true,linkcolor=blue,anchorcolor=blue,citecolor=blue,filecolor=blue,urlcolor=blue,bookmarksnumbered=true,pdfview=FitB}

\newcommand{\bk}{{{\bf{k}}}}

\newcommand{\br}{{{\bf{r}}}}
\newcommand{\bR}{{{\bf{R}}}}

\newcommand{\beqa}{\begin{eqnarray}}
\newcommand{\eeqa}{\end{eqnarray}}

\newcommand{\ua}{\uparrow}
\newcommand{\da}{\downarrow}

\newcommand{\bdelta}{{\boldsymbol \delta}}

\begin{document}

\hsize\textwidth\columnwidth\hsize\csname@twocolumnfalse\endcsname

\title{Spin-valley density wave in moir\'e materials}
\author{Constantin Schrade and Liang Fu}
\affiliation{Department of Physics, Massachusetts Institute of Technology, 77 Massachusetts Ave., Cambridge, MA 02139}
\date{\today}

\vskip1.5truecm
\begin{abstract}
We introduce and study a minimum two-orbital Hubbard model on a triangular lattice, which captures the key features of both the trilayer ABC-stacked graphene-boron nitride heterostructure and twisted transition metal dichalcogenides in a broad parameter range. Our model comprises first- and second-nearest neighbor hoppings with valley-contrasting flux that accounts for trigonal warping in the band structure. For the strong-coupling regime with one electron per site, we derive a spin-orbital exchange Hamiltonian and find the semiclassical ground state to be a spin-valley density wave. 
We show that a relatively small second-neighbor exchange interaction is sufficient to stabilize the ordered state against quantum fluctuations. Effects of spin- and valley Zeeman fields as well as thermal fluctuations are also examined.
\end{abstract}


\maketitle
Moir\'e materials are layered $2d$ crystals in which a lattice mismatch or a rotational misalignment gives rise to a long-period superlattice structure. These moir\'e superlattices host narrow mini-bands that promise enhanced correlation effects \cite{bib:Bistritzer2011,bib:Magaud2012}. Recent experiments have discovered correlated insulators, superconductivity, orbital ferromagnetism and spontaneous (quantum) Hall effect in several moir\'e materials including twisted bilayer graphene \cite{bib:Cao20181,bib:Cao20182,bib:Kerelsky2018,bib:Choi2019,bib:Yankowitz2019,bib:Cao2019,bib:Sharpe2019}, trilayer ABC-stacked graphene (TG) on hexagonal boron nitride (h-BN) \cite{bib:Chen2018,bib:Chen2019,bib:Chen20192}, and twisted transition metal dichalcogenides (TMDs) \cite{bib:Shih2019,bib:Jauregui2019}.

A paradigmatic approach for studying such correlated electron phenomena is the Hubbard model. For the aforementioned moir\'e materials, the effective Hubbard model comprises both spin and orbital degrees of freedom \cite{bib:Xu2018,bib:Xu2019,bib:Yuan2018,bib:Koshino2018} arising from the $K,K'$-valleys of the original Brillouin zone. Since the separation of $K,K'$-valleys is much larger than the reciprocal vector of the moire superlattice, inter-valley  hybridization is weak, thus, leading to Hubbard models with emergent symmetries.

A first concrete example are AB-stacked bilayers of TMDs which at small twist angle form a triangular superlattice \cite{bib:Wu2018,bib:Wu2019}. A recent work \cite{bib:Wu2019} has found that the topmost moir\'e valence bands of this material can be described by a two-orbital Hubbard model where each orbital resides in one of the two layers and electron's spin is locked to the valley. When the small layer separation is neglected, intra- and interlayer Coulomb repulsions are equal, which yields an interaction with SU(4)-symmetry.

A second example is TG/h-BN \cite{bib:Koshino2009,bib:Po2018,bib:Zhu20181,bib:Zhu20182,bib:Zhang2018,bib:Chittari2019,bib:Zhang2019,bib:Classen2019}. In this heterostructure, a vertical electric field enables a high degree of band structure tunability and permits the realization of a two-orbital Hubbard model on a triangular lattice \cite{bib:Xu2018,bib:Xu2019} with valley-contrasting flux \cite{bib:Po2018,bib:Zhu20181,bib:Zhu20182,bib:Zhang2018,bib:Chittari2019,bib:Zhang2019,bib:Classen2019}. This flux breaks SU(4)-symmetry while preserving charge and spin conservation within each valley.

A last example is twisted bilayer graphene where two graphene sheets are stacked with a small twist angle.
Theoretical works have constructed manifestly-symmetric, maximally-localized Wannier orbitals \cite{bib:Koshino2018,bib:Kang2018} and derived a two-orbital Hubbard model on a honeycomb lattice with extended interactions \cite{bib:Koshino2018,bib:Guinea2018,bib:Kang2018}. 
In both TG/h-BN and twisted bilayer graphene, the two orbitals in the effective Hubbard model correspond to Wannier states from the $K, K'$-valleys.

In this work, we introduce and study a minimum two-orbital Hubbard model on a triangular lattice, which captures key features of both TG/h-BN and twisted TMDs in a broad parameter range.
Our model includes first- and second-neighbor (NN) hopping as well as on-site interaction $U$. The first-NN hopping is complex and has opposite phases for the two valleys accounting for a valley-contrasting flux, while the second-NN hopping is real due to crystal symmetry. Focusing on the large-$U$ limit with one electron per site, we derive a spin-orbital exchange Hamiltonian $H_J$ with $\text{SU(2)}\times\text{SU(2)}\times\text{U(1)}$-symmetry, associated with spin and charge conservation within each valley.  By solving $H_J$ in the semiclassical limit, we find a ``spin-valley density wave" ground state with four-sublattice order. We also show by a spin-wave analysis that a relatively small second-neighbor exchange interaction is sufficient to stabilize the order against quantum fluctuations at zero temperature. We show that thermal melting of the $T=0$ ground state restores spin rotation symmetry and may lead to a valley density wave state at low temperature, which breaks discrete lattice and time-reversal symmetries. Finally, we examine the effects of spin and valley Zeeman fields and discuss experimental signatures of the predicted density wave states in TG/h-BN and twisted TMD.

\textit{Model.}
We begin with a detailed description of our proposed Hubbard model for TG/hBN and twisted TMDs.

\begin{figure}[!b] \centering
\includegraphics[width=1\linewidth] {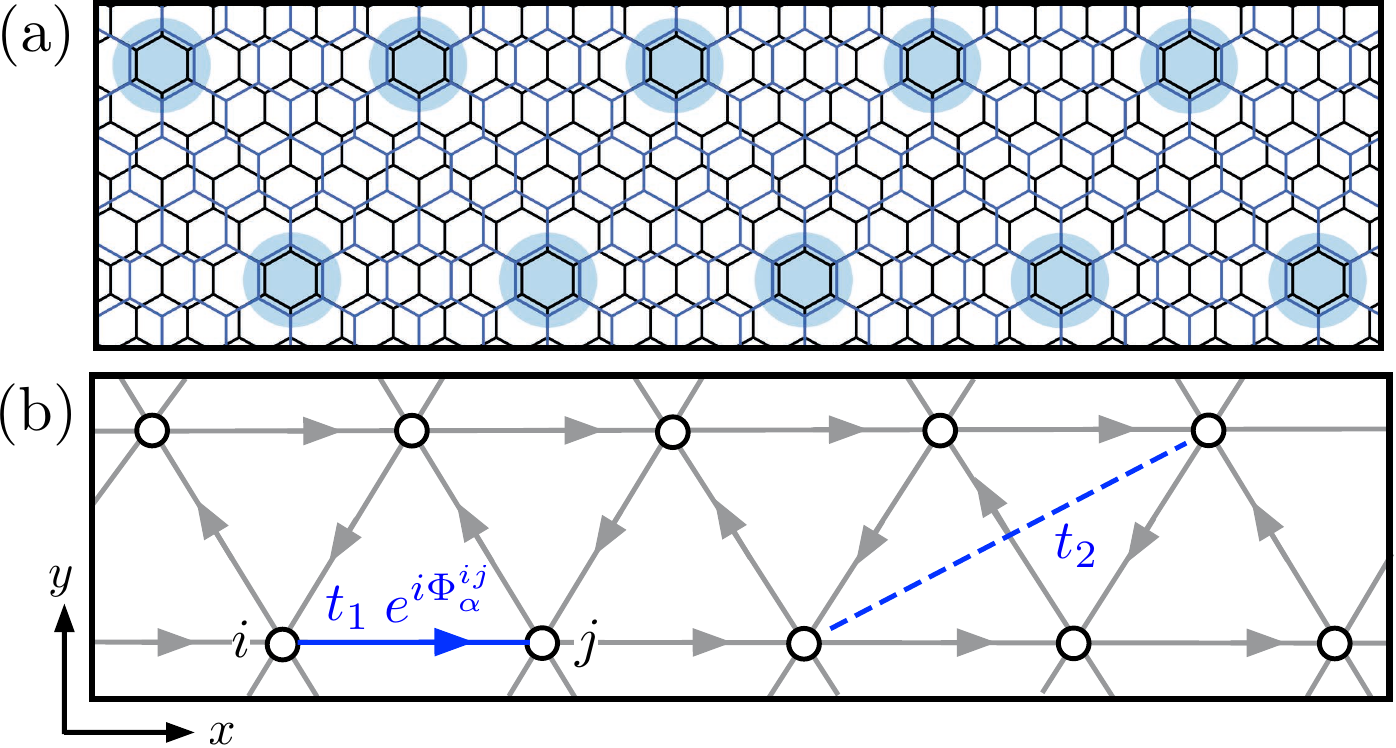}
\caption{(Color online)
(a) Schematic plot of the triangular moir\'e superlattice formed by TG (black) and h-BN (blue). 
(b) Triangular lattice with directed bonds. First-NN hoppings along the bond direction acquire a valley-contrasting
phase, $t_{1}e^{\Phi^{ij}_{\alpha}}$. Second-NN hoppings $t_{2}$ are real-valued.
}\label{fig:1}
\end{figure}

We will first consider TG/h-BN. In this heterostructure, both individual components, TG and h-BN, have a $1.5\%$-mismatch of lattice constants which results in a triangular moir\'e superlattice, see Fig.~\ref{fig:1}(a). For this superlattice, the microscopic symmetries are three-fold rotations $C_{3}$ around the axis perpendicular to the TG/h-BN sheets, mirror reflection symmetry $M$, and time-reversal symmetry.

The mini-band structure in TG/h-BN arises from the moir\'e potential of h-BN acting on  low-energy electrons in TG. \cite{bib:Koshino2009,bib:Zhu20181,bib:Zhu20182}.  An important experimental parameter for tuning the bandwidth and topology of the mini-bands is the external electric field \cite{bib:Chen2018,bib:Chen2019} that provides a potential difference between the top and bottom graphene sheets. In TG/hBN, depending on the the sign of the potential difference, the mini-band structure is either in a ``Hubbard regime" with zero Chern number \cite{bib:Po2018,bib:Zhu20181,bib:Zhu20182,bib:Zhang2018,bib:Classen2019} or in a ``Quantum Hall regime" with finite valley Chern number \cite{bib:Zhang2019,bib:Chittari2019,bib:Chen20192,bib:Song2015}. In this work, we focus on the ``Hubbard regime" without Chern number.

The Hubbard regime is realized when electrons in TG are pulled towards h-BN by the external electric field. 
The resulting mini-band structure can be intuitively understood from the deep potential limit, where each minimum of the moir\'e potential creates a localized Wannier orbital. Since the potential minima form a triangle lattice, one naturally expects a triangle-lattice tight-binding model for TG/h-BN, as demonstrated by previous band structure calculations \cite{bib:Chittari2019,bib:Yuan2019}. Since the hopping matrix elements decay rapidly with the distance, we study a \textit{minimal} model for TG/h-BN that only retains the dominant hopping terms. The full Hamiltonian of our model is $H=H_{0}+H_{I}$, where the single-particle Hamiltonian $H_0$ is given by,
\begin{equation}
\hspace{-3pt}
H_{0}
=\sum_{\alpha}
\left(t_{1}
\sum_{\langle i,j\rangle}e^{i\Phi^{ij}_{\alpha}}c^{\dag}_{i\alpha}c_{j\alpha}
+t_{2}
\sum_{\langle\langle i,j\rangle\rangle} c^{\dag}_{i\alpha}c_{j\alpha}
+
...
\right),
\end{equation}
where $c_{i\alpha}$ annihilates an electron  at site $i$ in state $\alpha=(\sigma,\tau)$ with spin $\sigma=\ua,\da$ and orbital $\tau=\pm$ associated with the $K,K'$-valleys. $t_{1}$, $t_{2}$ are the dominant hopping amplitudes between first and second-NN sites. The $t_{1}$-hoppings are generally complex and carry phases $\Phi^{ij}_{\alpha}=-\Phi^{ji}_{\alpha}$, which are independent of spin and opposite for the two valleys. Microscopically, these phases arises from the trigonally warped Dirac dispersions at the $K,K'$-valleys and are allowed by symmetry.
The total flux piercing through each elementary triangle is $3\Phi_{\alpha}\equiv\Phi^{ij}_{\alpha}+\Phi^{jk}_{\alpha}+\Phi^{ki}_{\alpha}$ where $\Phi_{\sigma,+}=-\Phi_{\sigma,-}\equiv\Phi$ and $i,j,k$ are three consecutive triangle sites along the directed first-NN bonds, see Fig.~\ref{fig:1}(b). In comparison, the $t_{2}$-hoppings are real-valued due to the combination of reflection $x\rightarrow -x$ and time-reversal symmetry, which acts within each valley. In this work, the $t_{2}$ hopping will play an important role as shown below.

Second, the dominant term in the projected Coulomb interactions onto the narrow mini-bands is the on-site density interaction,
\begin{equation}
\begin{split}
\label{HI}
H_{I}&=\frac{U}{2}\sum_{i}(n_{i}-n_{0})^{2},
\end{split}
\end{equation}
where $n_{i}=\sum_{\sigma,\xi}c^{\dag}_{i\sigma\xi}c_{i\sigma\xi}$ is the total number electrons on the $i$-site, $n_{0}$ controls the filling and $U$ is the interaction amplitude. In this work, we will focus on the regime where kinetic exchange due to single-particle hopping dominates over direct interactions between electrons on different sites. We remark that previous works have considered alternative mechanism of SU(4) symmetry breaking due to extended interactions instead of valley-dependent single-particle hopping \cite{bib:Zhang2018,bib:Xu2019}.

\textit{Variational study.}
We now proceed to study of our two-orbital Hubbard model in the strong-coupling limit,  $U\gg t_1,t_2$. Such an approach is complimentary to previous studies in the weak coupling limit \cite{bib:Isobe2018,bib:Liu2018,bib:Laksono2018,bib:Sherkunov2018,bib:Kennes2018,bib:Lin2018,bib:You2018,bib:Gonzlez2019,bib:Guinea2018} as well as to numerical works \cite{bib:Xu20182,bib:Zhu2018,bib:Classen2019}. 
Specifically, we will focus on the filling of one electron per site and carry out a perturbative expansion to second order in $t_1,t_2$ leading to a spin-orbital exchange interaction,
\begin{equation}
H_{J}=\sum_{\alpha,\beta}\left(J_{1}\sum_{\langle i,j \rangle}
e^{i(\Phi^{ij}_{\beta}-\Phi^{ij}_{\alpha})}
T^{\alpha}_{\beta,i}T^{\beta}_{\alpha,j}
+
J_{2}\sum_{\langle\langle i,j \rangle\rangle}
T^{\alpha}_{\beta,i}T^{\beta}_{\alpha,j}
\right).
\label{Heff}
\end{equation}
Here, $J_{1}=2t_{1}^{2}/U$, $J_{2}=2t_{2}^{2}/U$ are antiferromagnetic exchange couplings and $T^{\alpha}_{\beta}=|\beta\rangle\langle\alpha|$ are SU(4) generators that act on the spin-orbital basis states $\left|+,\ua\right\rangle$,
$\left|+,\da\right\rangle$, $\left|-,\ua\right\rangle$,
$\left|-,\da\right\rangle$. The SU(4) generators satisfy $\sum_{\alpha}T^{\alpha}_{\alpha}=1$, $(T^{\alpha}_{\beta})^{\dag}=T^{\beta}_{\alpha}$, $[T^{\alpha}_{\beta},T^{\beta'}_{\alpha'}]=\delta_{\alpha\alpha'}T^{\beta'}_{\beta}-\delta_{\beta\beta'}T^{\alpha}_{\alpha'}$. Despite being written in terms of SU(4) generators, $H_{J}$ is not SU(4) symmetric when $\Phi^{ij}_{\alpha}\neq0$, as the exchange of electrons in different orbitals picks up a orbital-dependent phase factor. This phase factor shows up in the first term $\propto J_{1}$ and leads to a breaking of SU(4) down to SU(2)$\times$SU(2)$\times$U(1) with generators $\vec{\sigma}\oplus I$, $I \oplus \vec{\sigma}$, and $I \oplus (-I)$, where $\sigma$'s are $2\times 2$ Pauli matrices associated with spin and $\oplus$ denotes direct sum of the two valleys. 

Next, we will determine the ground states of $H_{J}$ in the semiclassical approximation.
For this purpose, we consider the following product state,
\begin{equation}
\left|\Psi\right\rangle=\prod_{i}
\left(
\sum_{\alpha}
v_{\alpha,i}\left| \alpha \right\rangle_{i}\right),
\end{equation}
where we have defined complex and normalized vectors $\boldsymbol{v}_{i}=(v_{1,i},v_{2,i},v_{3,i},v_{4,i})^{T}$ for each site.
To find the variational ground states based on this ansatz, we note that the two terms in the effective Hamiltonian permute the states on first-NN and second-NN sites, $T^{\alpha}_{\beta,i}T^{\beta}_{\alpha,j}=|\beta_{i},\alpha_{j}\rangle\langle\alpha_{i},\beta_{j}|$. Hence, the variational ground states need to minimize \cite{bib:supplemental},
\begin{align}
\label{Min}
\langle\Psi|H_{J}|\Psi\rangle
&=
J_{1}
\sum_{\langle i,j \rangle} \left|
\sum_{\alpha}
e^{i\Phi^{ij}_{\alpha}}v^{*}_{\alpha,i}v_{\alpha,j}
\right|^{2}
+
J_{2}
\sum_{\langle\langle i,j \rangle\rangle} |\boldsymbol{v}^{*}_{i}\cdot \boldsymbol{v}_{j}|^{2},
\end{align}
Since $J_{1},J_{2}>0$, the energy of each second-NN bond is minimized when $\boldsymbol{v}_{i}$, $\boldsymbol{v}_{j}$ are orthogonal and it is minimized for each first-NN bond when $|
\sum_{\alpha}
e^{i\Phi^{ij}_{\alpha}}v^{*}_{\alpha,i}v_{\alpha,j}|=0$. Notably, we find that for all the first- and second-NN bonds and all values of valley-contrasting flux the conditions are satisfied by the spin-valley density wave ground state shown in Fig.~\ref{fig:2}. Three remarks are in order:

\begin{figure}[!b] \centering
\includegraphics[width=1\linewidth] {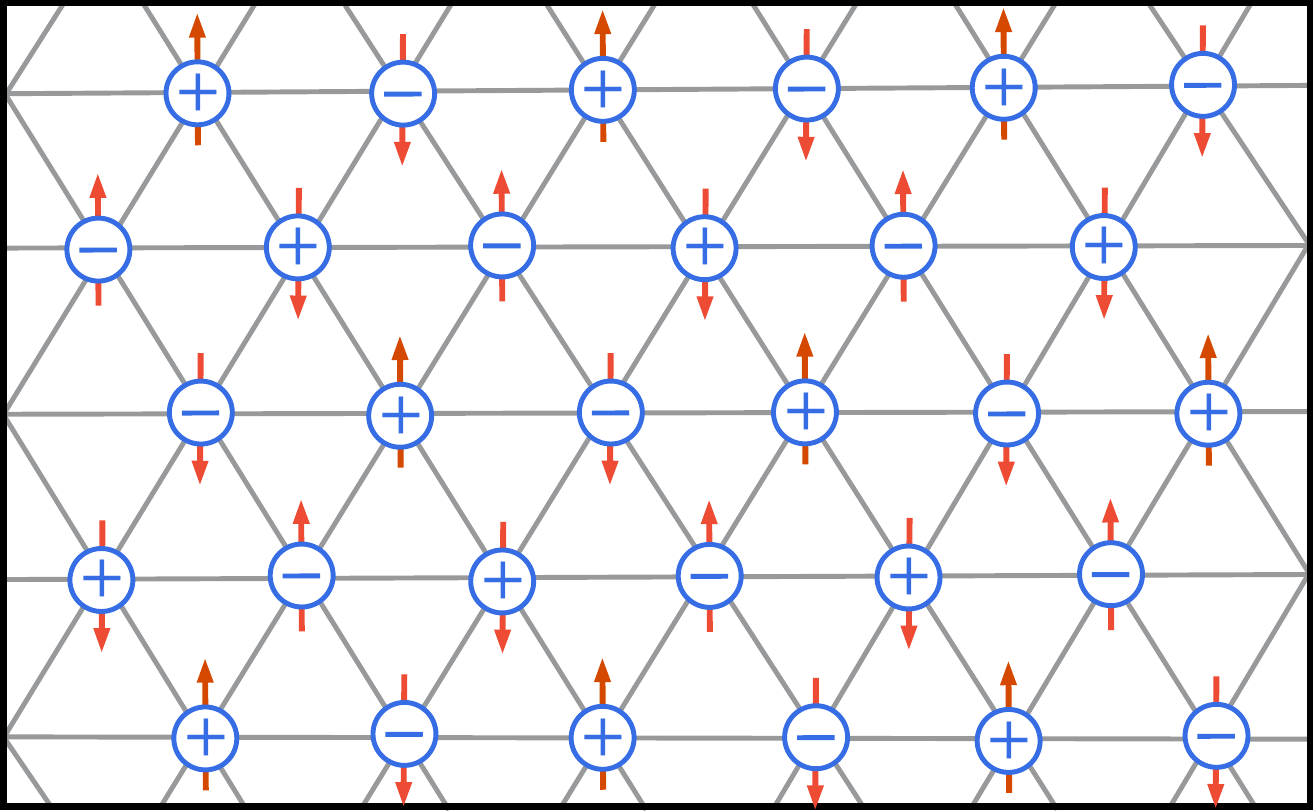}
\caption{(Color online)
Spin-valley density wave with a four-sublattice order which is a variational ground state for all values of
valley-contrasting flux $\Phi_{\alpha}$. The spin $\ua,\da$-states are shown in red and the orbital $\pm$-states are shown in blue.
}\label{fig:2}
\end{figure}

(1) The ground state for our two-orbital model exhibits \textit{four}-sublattice spin-valley density wave order, a triplet-$Q$ state with the commensurate wavevector $\Gamma M$. The situation is thus markedly different from the SU(2) Heisenberg model on the triangular lattice \cite{bib:Zhu2015,bib:Hu2015,bib:Gong2017} for which the ground state for small  $J_{2}/J_{1}$ is the $120^{\circ}$-state with \textit{three}-sublattice order at the wavevector $\Gamma K$.

(2) The presence of the valley-contrasting flux affects the ground state manifold: If $\Phi^{ij}_{\alpha}=0$, all semiclassical ground states have mutually orthogonal states on first-NN and second-NN bonds and can be generated from the configuration in Fig.~\ref{fig:2} by a global SU(4) rotation \cite{bib:Penc2003}. If $\Phi^{ij}_{\alpha}\neq0$ only a subset of these states, generated from the configuration in Fig.~\ref{fig:2} by SU(2)$\times$SU(2) rotations, are semiclassical ground states. For example, $\boldsymbol{v}_{i}=(1,0,1,0)^{T}$ and $\boldsymbol{v}_{j}=(1,0,-1,0)^{T}$ do not minimize the semiclassical energy of an first-NN bond if $\Phi^{ij}_{\alpha}\neq0$ despite being mutually orthogonal. The manifold of ground state we found for $\Phi^{ij}_{\alpha}\neq0$ is parameterized by two {\it independent} unit vectors denoting the spin axis associated with each valley.

(3) In addition to breaking the spin rotation symmetry from SU(2)$\times$SU(2) to U(1)$\times$U(1), the spin-valley density wave state breaks lattice translation symmetry.  However, it preserves the U(1) valley number symmetry and, in particular, is valley-unpolarized. Our ground state has a finite energy gap to all excitations with an unbalanced occupation of the two valleys.

\textit{Quantum fluctuations.}
To understand the stability for the spin-valley density wave ground state of Fig.~\ref{fig:2}, we now proceed by
studying the effects of quantum fluctuations with a generalized Holstein-Primakoff (HP) transformation \cite{bib:Papanicolaou1984,bib:Papanicolaou1988,bib:Joshi1999}. We, therefore, assign the $\alpha$-spin-orbital basis state to each site of the $\Lambda_{\alpha}$-sublattice. Based on this choice, the generalized HP transformation for a site $i\in\Lambda_{\alpha}$ is given by $T^{\alpha}_{\alpha,i}=M-\sum_{\beta\neq\alpha}b^{\alpha\dag}_{\beta,i}b^{\alpha}_{\beta,i}$, $T^{\alpha}_{\beta,i}=b^{\alpha\dag}_{\beta,i}(M-\sum_{\beta\neq\alpha}b^{\alpha\dag}_{\beta,i}b^{\alpha}_{\beta,i})^{1/2}$, $T^{\beta}_{\alpha,i}=(M-\sum_{\beta\neq\alpha}b^{\alpha\dag}_{\beta,i}b^{\alpha}_{\beta,i})^{1/2}b^{\alpha}_{\beta,i}$ and $T^{\beta'}_{\beta,i}=b^{\alpha\dag}_{\beta,i}b^{\alpha}_{\beta',i}$ where $b^{\alpha}_{\beta,i}$ denote bosonic operators with $\beta\neq\alpha$ and $M$ is a positive integer. Next, we insert the HP transformation in the effective Hamiltonian of Eq.~\eqref{Heff}, perform a $1/M$-expansion, and only retain terms that are quadratic in the bosonic operators. The exchange interaction then takes on the form $H_{J}\approx M\sum_{\alpha\neq\beta}H_{\alpha\beta}$ where
\begin{equation}
\label{Hab}
H_{\alpha\beta}=J_{1}\sum_{
\mathclap{\substack{
\langle i,j\rangle\\
i\in\Lambda_{\alpha},j\in\Lambda_{\beta}
                  }}
                  }
                  A^{\dag}_{ij}A_{ij}
                  +
J_{2}\sum_{
\mathclap{\substack{
\langle\langle i,j\rangle\rangle\\
i\in\Lambda_{\alpha},j\in\Lambda_{\beta}
                  }}
                  }
                  B^{\dag}_{ij}B_{ij},
                  \end{equation}
and we have introduced the bond-operators
$
A^{\dag}_{ij}=e^{i\Phi^{ij}_{\beta}}b^{\beta\dag}_{\alpha,j}+e^{i\Phi^{ij}_{\alpha}}b^{\alpha}_{\beta,i}
$
and
$
B^{\dag}_{ij}=b^{\beta\dag}_{\alpha,j}+b^{\alpha}_{\beta,i}
$
In this representation of the effective Hamiltonian, $b^{\beta}_{\alpha}$ only pairs with $b^{\alpha}_{\beta}$ which implies that that individual $H_{\alpha\beta}$-terms
decouple and can, thus, be studied independently of each other.
For deriving the aforementioned stability phase diagram, we proceed in two steps:

First, we Fourier transform the Hamiltonian of Eq.~\eqref{Hab} to momentum space and
diagonalize it by means of a Bogoliubov transformation. This gives the dispersions,
\begin{equation}
\label{Dispersion}
\omega^{\alpha\beta}_{\bk}=2(J_{1}+J_{2})
\sqrt{1-|\gamma^{\alpha\beta}_{\bk}|^{2}},
\end{equation}
where $\bk$ is a momentum in the reduced Brillouin zone (RBZ) of the four-sublattice ordered spin-valley density wave state. Moreover, we defined the factor $\gamma^{\alpha\beta}_{\bk}=[J_{1}\cos(\bk\cdot\bR^{(1)}_{\alpha\beta}+\Phi_{\alpha}-\Phi_{\beta})+
J_{2}\cos(\bk\cdot\bR^{(2)}_{\alpha\beta})]/(J_{1}+J_{2})$. Here, $\bR^{(1)}_{\alpha\beta}$ is a vector that connects the first-NN  sites of the $\Lambda_{\alpha}$- and $\Lambda_{\beta}$-sublattices and points along the bond direction. Similarly, $\bR^{(2)}_{\alpha\beta}$ is a vector connects the second-NN sites of the $\Lambda_{\alpha}$- and $\Lambda_{\beta}$-sublattices. At this point, two comments are in order:

(1) If $J_{2}=0$, the dispersions vanish along the line $\bk\cdot\bR^{(1)}_{\alpha\beta}+\Phi_{\alpha}-\Phi_{\beta}=0$ and we anticipate that the resulting low-energy quantum fluctuations destroy the spin-valley density wave order.
This means that the spin-valley density wave order for TG/h-BN is not possible in previous models with $t_{2}=0$ \cite{bib:Po2018,bib:Zhu20181,bib:Zhu20182,bib:Classen2019}.

(2) If $J_{2}\neq0$, the dispersion vanish at discrete points in the RBZ. We expect that this behavior will reduce low-energy quantum fluctuations and will be crucial for stabilizing the spin-valley density wave order.

To confirm these arguments, we compute the reduction of the $\alpha$-ordered moment due to the quantum fluctuations,
$\langle T^{\alpha}_{\alpha,i} \rangle=M-\langle \sum_{\beta\neq\alpha} b^{\alpha\dag}_{\beta,i}b^{\alpha}_{\beta,i}\rangle$
where the $i$-site is on the $\Lambda_{\alpha}$-sublattice. We find that in momentum space,
\begin{equation}
\label{Moment}
\hspace{-6.5pt}
\langle T^{\alpha}_{\alpha,i} \rangle
=
M-\frac{1}{2}
\sum_{\beta\neq\alpha}
\left\langle
\frac{1}{\sqrt{1-|\gamma^{\alpha\beta}_{\bk\tau}|^{2}}}
-1
\right\rangle_{\text{RBZ}}.
\end{equation}
Here, $\langle...\rangle_{\text{RBZ}}$ denotes the average over the RBZ. By numerically evaluating Eq.~\eqref{Moment} and setting $M=1$, we find that $\langle T^{\alpha}_{\alpha,i}\rangle>0$ for
$J_{2}/J_{1}\gtrapprox 0.12$ \cite{bib:supplemental}. This threshold does not depend on the orbital-contrasting flux as the latter only provides a constant momentum-space displacement in the dispersion of Eq.~\eqref{Dispersion} and, thereby, does not change the RBZ-average. Accordingly, our prediction is that the system transitions from a disordered phase for $J_{2}/J_{1}\lessapprox 0.12$ to a phase with a stable spin-valley density wave order for $J_{2}/J_{1}\gtrapprox 0.12$. The nature of the disordered phase is an interesting question we leave to a separate study.

\textit{Zeeman field effects.}
We will now study the effects of spin/orbital-Zeeman fields in our spin-orbital model,
\begin{equation}
\label{Hfield}
H =  H_{J} - h_{\sigma} \sum_{i} \sigma^{z}_{i} - h_{\tau}   \sum_{i} \tau^{z}_{i},
\end{equation}
where  $\sigma$ and $\tau$ are $2\times 2$ Pauli matrices acting in spin and orbital subspace respectively.  
In TG/h-BN, the spin-Zeeman field can be realized by an in-plane magnetic field and the valley-Zeeman field by an out-of-plane magnetic field.

First, we set the orbital-Zeeman field to zero, $h_{\tau}=0$, and consider the case of a spin-Zeeman field $h_{\sigma}$.
A large $h_{\sigma}$ freezes the spin degrees of freedom and we can recast $H_{J}$ into a form that includes only the remaining orbital degrees of freedoms. Neglecting  the small $J_2$ term, we find that,
\begin{equation}
\label{StrongField}
H \approx
J_{1}
\sum_{\langle i,j \rangle}
[1+\boldsymbol{\tau}_{i}\cdot
\boldsymbol{\Omega}_{z}(2\Phi^{ij})
\cdot \boldsymbol{\tau}_{j}
]/2.
\end{equation}
Here, $\boldsymbol{\Omega}_{z}(2\Phi)$ is a rotation matrix about the $z$-axis by a $2\Phi$-angle. Eq.~\eqref{StrongField} can also be written as an anisotropic exchange interaction with a Dzyaloshinskii-Moriya term,
$
\sim
J_{1}[1+\tau^{z}_{i}\tau^{z}_{j}+\cos(2\Phi^{ij})(\tau^{x}_{i}\tau^{x}_{j}+\tau^{y}_{i}\tau^{y}_{j})
+
\sin(2\Phi^{ij})(\tau^{x}_{i}\tau^{y}_{j}-\tau^{y}_{i}\tau^{x}_{j})]/2
$ with $\Phi^{ij}=-\Phi^{ji}=\Phi$.

To find the semiclassical ground state of Eq.~\eqref{StrongField}, we minimize the expectation value of $H$ with respect to the orientation of orbital pseudospin $\vec \tau$ at every site.
Here, we will focus on a particular case when the lower energy bound $E_{ij}\geq J_{1}(1-S^{2})/2$ is saturated for all bonds. Such a situation is achieved for $\Phi=\pi/6$ and, in this case, we find that the unique ground state is the $120^{\circ}$ planar spin state, where spins lie on the $xy$ plane. Since the $120^{\circ}$-state has three-sublattice order distinct from the four-sublattice order of the spin-valley density wave state at zero Zeeman field, we predict that by increasing the
spin Zeeman field in TG/h-BN a phase transition between insulating states with different spin-valley density wave orders can be achieved.

Next, we set the spin-Zeeman field to zero, $h_{\sigma}=0$, and consider a  finite valley-Zeeman field $h_{\tau}$. Since our ground state preserves the valley $U(1)$ symmetry and has a gap to valley excitations, we expect that the ground state is unchanged by a small valley Zeeman field. However, for a strong valley-Zeeman field, the system can lower its energy by aligning orbital pseudospins in the same direction, thus, effectively freezing the orbital degrees of freedom. We are then left with a $J_{1}-J_{2}$ Heisenberg model of spins on a triangular lattice for which the three-sublattice ordered $120^{\circ}$-state is the semiclassical ground state when $J_{2}/J_{1}\ll 1$. 

For both spin and valley Zeeman fields, the spin-valley density wave state at zero/small field and the polarized state at high-field have distinct symmetries, and hence, must be separated by phase transitions. 

\textit{Thermal melting}. 
Finally, we discuss the effect of thermal fluctuations. Since the spin-valley density wave ground state breaks spin rotation symmetry, at finite temperature long-range order is destroyed by thermal fluctuations associated with Goldstone modes. However, a partially ordered state with composite order parameters that only break discrete symmetries may exist at low temperature. One such state is a unidirectional valley density wave (or valley stripe) at wavevector $\Gamma M$, in which spin order is restored but lattice translation symmetry is broken. 

\textit{Conclusion.}
We have introduced and studied a two-valley Hubbard model on a triangular lattice for describing the correlated insulator phases of TG/h-BN and twisted TMDs. Specifically, in the strong coupling limit, we have identified a four-sublattice ordered spin-valley density wave state as an ordered ground state that appears at moderate values of beyond-NN hoppings. Moreover, we have demonstrated that this spin-valley density wave-state undergoes a phase transition to a $120^{\circ}$-state in either spin- or orbital space upon increasing the magnitude of an external spin- or valley-Zeeman field.

\textit{Acknowledgments}
We would like to thank Zhen Bi, Noah F. Q. Yuan and Hiroki Isobe for helpful discussions. We also thank Feng Wang and Abhay Pasupathy for simulating discussions on trilayer graphene and twisted bilayer TMD respectively.
This work was supported by DOE Office of Basic Energy Sciences, Division of Materials Sciences and Engineering under Award $\text{DE-SC0018945}$. LF was supported in part by a Simons Investigator Award from the Simons Foundation.

\begin{widetext}

\newpage

\onecolumngrid

\bigskip

\begin{center}
\large{\bf Supplemental Material to `Spin-valley density wave in moir\'e materials' \\}
\end{center}
\begin{center}
Constantin Schrade and Liang Fu
\\
{\it Department of Physics, Massachusetts Institute of Technology, 77 Massachusetts Ave., Cambridge, MA 02139}
\end{center}

In the Supplemental Material, we provide more details on the derivation of the exchange interaction, the semiclassical bond energy, and the spin-wave dispersions.

\section{Exchange interaction}
In this first section of the Supplemental Material, we will derive the effective exchange interaction between first and second-nearest sites as given in Eq.~(3) of the main text. As a starting point, we note that to second order in $t_{1},t_{2}$, the general form of the effective Hamiltonian is given by
\begin{equation}
H_{J}=-\frac{PH_{0}(1-P)H_{0}P}{H_{I}-E_{0}},
\label{Heffgeneral}
\end{equation}
Here, we have introduced the operator $P$ that projects on the ground states at energy $E_{0}$ with all sites singly-occupied.

Next, we evaluate the effective Hamiltonian in Eq.~\eqref{Heffgeneral} by computing all possible sequences of intermediate states, see Fig.~\ref{fig:sm1}.
More specifically, for a fixed $\langle i,j\rangle$-bond, an electron can hop from the $j$-site to the $i$-site and back,
\begin{equation}
\sum_{\alpha,\beta}
P(e^{-i\Phi^{ij}_{\alpha}}c^{\dag}_{j\alpha}c_{i\alpha})(e^{i\Phi^{ij}_{\beta}}c^{\dag}_{i\beta}c_{j\beta})P
=
\sum_{\alpha,\beta}
[
1
-
e^{i(\Phi^{ij}_{\beta}-\Phi^{ij}_{\alpha})}
P(c^{\dag}_{i\beta}c_{i\alpha}c^{\dag}_{j\alpha}c_{j\beta})P
]
=
\sum_{\alpha,\beta}
[
1
-
P(e^{i(\Phi^{ij}_{\beta}-\Phi^{ij}_{\alpha})}T^{\alpha}_{\beta,i}T^{\beta}_{\alpha,j})P
]
.
\label{sequence1}
\end{equation}
Alternatively, the electron can also hop from the $i$-site to the $j$-site and back,
\begin{equation}
\sum_{\alpha,\beta}
P(e^{i\Phi^{ij}_{\beta}}c^{\dag}_{i\beta}c_{j\beta})(e^{-i\Phi^{ij}_{\alpha}}c^{\dag}_{j\alpha}c_{i\alpha})P
=
\sum_{\alpha,\beta}
[
1
-
e^{i(\Phi^{ij}_{\beta}-\Phi^{ij}_{\alpha})}
P(c^{\dag}_{i\beta}c_{i\alpha}c^{\dag}_{j\alpha}c_{j\beta})P
]
=
\sum_{\alpha,\beta}
[
1
-
P(e^{i(\Phi^{ij}_{\beta}-\Phi^{ij}_{\alpha})}T^{\alpha}_{\beta,i}T^{\beta}_{\alpha,j})P
]
.
\end{equation}
If we combine the two types of sequences, multiply by the appropriate energy denominator, and repeat these same steps for the $\langle\langle i,j\rangle\rangle$-bonds, we arrive at the effective exchange interaction,
\begin{equation}
H_{J}=J_{1}\sum_{\langle i,j \rangle}\sum_{\alpha,\beta}
e^{i(\Phi^{ij}_{\beta}-\Phi^{ij}_{\alpha})}
T^{\alpha}_{\beta,i}T^{\beta}_{\alpha,j}
+
J_{2}\sum_{\langle\langle i,j \rangle\rangle}\sum_{\alpha,\beta}
T^{\alpha}_{\beta,i}T^{\beta}_{\alpha,j}.
\end{equation}
Because of the $\alpha,\beta$-summation, we note that this expression for $H_{J}$ is invariant under a swapping the $\langle i,j\rangle$-bond indices or, equivalently, reversing the $\langle i,j\rangle$-bond direction. In particular, this means that the $\langle i,j\rangle$-summation in $H_{J}$ does not require us to consider directed bonds as in the case of $H_{0}$.

\section{Minimization condition}
In this second section of the Supplemental Material, we provide more details on the derivation of the minimization condition given in Eq.~(5) of the main text. More specifically, we will focus on deriving the first term $\propto J_{1}$.

As a first step, we consider a fixed $\langle i,j\rangle$-bond and notice that the action of $H_{J}$ on this bond is given by
\begin{equation}
H_{ij}\equiv
\sum_{\alpha,\beta}
e^{i(\Phi^{ij}_{\beta}-\Phi^{ij}_{\alpha})}
T^{\alpha}_{\beta,i}T^{\beta}_{\alpha,j}=
\sum_{\alpha,\beta}
e^{i(\Phi^{ij}_{\beta}-\Phi^{ij}_{\alpha})}
|\beta_{i},\alpha_{j}\rangle\langle\alpha_{i},\beta_{j}|.
\end{equation}
As a second step, we consider a state of the $\langle i,j\rangle$-bond we will assume to be product state in bond space,
\begin{equation}
\boldsymbol{v}_{ij}=
\left(\sum_{\alpha}v_{\alpha,i}\left| \alpha \right\rangle_{i}\right)
\left(\sum_{\beta}v_{\beta,j}\left| \beta \right\rangle_{j}\right)
=
\sum_{\alpha,\beta}
v_{\alpha,i}\hspace{2pt}v_{\beta,j}
\left| \alpha_{i}, \beta_{j} \right\rangle
\end{equation}
As a final step, we evaluate the expression,
\begin{equation}
\begin{split}
\boldsymbol{v}^{\dag}_{ij}\cdot H_{ij}\cdot \boldsymbol{v}_{ij}
&=
\sum_{\alpha,\beta}
\sum_{\alpha',\beta'}
\sum_{\alpha'',\beta''}
e^{i(\Phi^{ij}_{\beta''}-\Phi^{ij}_{\alpha''})}
v^{*}_{\alpha',i}\hspace{2pt}v^{*}_{\beta',j}
v_{\alpha,i}\hspace{2pt}v_{\beta,j}
\left\langle \alpha'_{i}, \beta'_{j} \right|\beta''_{i},\alpha''_{j}\rangle\langle\alpha''_{i},\beta''_{j}
\left| \alpha_{i}, \beta_{j} \right\rangle
\\
&=
\sum_{\alpha,\beta}
\sum_{\alpha',\beta'}
\sum_{\alpha'',\beta''}
e^{i(\Phi^{ij}_{\beta''}-\Phi^{ij}_{\alpha''})}\
v^{*}_{\alpha',i}\hspace{2pt}v^{*}_{\beta',j}
v_{\alpha,i}\hspace{2pt}v_{\beta,j}\
\delta_{\alpha'\beta''}
\delta_{\beta'\alpha''}
\delta_{\alpha'',\alpha}
\delta_{\beta'',\beta}
\\
&=
\sum_{\alpha,\beta}
e^{i(\Phi^{ij}_{\beta}-\Phi^{ij}_{\alpha})}\
v^{*}_{\beta,i}\hspace{2pt}v^{*}_{\alpha,j}
v_{\alpha,i}\hspace{2pt}v_{\beta,j}\
\\&=
\left(
\sum_{\alpha}
e^{-i\Phi^{ij}_{\alpha}}v_{\alpha,i}v^{*}_{\alpha,j}
\right)
\left(
\sum_{\beta}
e^{i\Phi^{ij}_{\beta}}v^{*}_{\beta,i}v_{\beta,j}
\right)
\\
&=
\left|
\sum_{\alpha}
e^{i\Phi^{ij}_{\alpha}}v^{*}_{\alpha,i}v_{\alpha,j}
\right|^{2}.
\end{split}
\end{equation}
This result corresponds to the first term $\propto J_{1}$ in Eq.~(5) of the main text.

\section{Dispersion relations}
In this third section of the Supplemental Material, we consider in more detail the derivation of the
dispersion relations given by Eq.~(7) in the main text. For clarity, we will initially set $J_{2}=0$ in our derivation.

First, we perform a generalized Holstein-Primakoff transformation as described in the main text.
If we only retain terms that are quadratic in the bosonic operators, we find that $H_{J}\approx M\sum_{\alpha\neq\beta}H_{\alpha\beta}$ with,
\begin{equation}
\begin{split}
H_{\alpha\beta}&=J_{1}\sum_{
\mathclap{\substack{
\langle i,j\rangle\\
i\in\Lambda_{\alpha},j\in\Lambda_{\beta}
                  }}
                  }
 b^{\alpha\dagger}_{\beta,i} b^{\alpha}_{\beta,i}
 + b^{\beta\dagger}_{\alpha,j} b^{\beta}_{\alpha,j}
 + e^{i(\Phi^{ij}_{\beta}-\Phi^{ij}_{\alpha})} b^{\alpha\dagger}_{\beta,i} b^{\beta\dagger}_{\alpha,j}
 + e^{-i(\Phi^{ij}_{\beta}-\Phi^{ij}_{\alpha})} b^{\alpha}_{\beta,i} b^{\beta}_{\alpha,j}.
 \end{split}
\end{equation}
In the following considerations, we define $\bdelta_{j}$ ($j=1,...,3$) to be lattice basis vectors pointing along the directed bonds of the triangular lattice and, as a result of rotational symmetry, we have $e^{i\Phi^{\br,\br+\bdelta_{j}}_{\alpha}}=e^{i\Phi_{\alpha}}$.

Second, we consider sites $\br\in\Lambda_{\alpha}$ with nearest-neighbors  $\br\pm\boldsymbol{\delta} \in \Lambda_{\beta}$. Then we can rewrite $H_{\alpha\beta}$ as,
\begin{equation}
\begin{split}
H_{\alpha\beta}&=
J_{1}
\sum_{\br}
\left[
 2b^{\alpha\dagger}_{\beta,\br} b^{\alpha}_{\beta,\br}
 +
 b^{\beta\dagger}_{\alpha,\br-\bdelta} b^{\beta}_{\alpha,\br-\bdelta}
  +
 b^{\beta\dagger}_{\alpha,\br+\bdelta} b^{\beta}_{\alpha,\br+\bdelta}
 \right]
 \\
 &+
 J_{1}
 \sum_{\br}
 \left[
  e^{i(\Phi^{\br,\br-\delta}_{\beta}-\Phi^{\br,\br-\bdelta}_{\alpha})}
  b^{\alpha\dagger}_{\beta,\br} b^{\beta\dagger}_{\alpha,\br-\bdelta}
  +
    e^{i(\Phi^{\br,\br+\bdelta}_{\beta}-\Phi^{\br,\br+\bdelta}_{\alpha})}
  b^{\alpha\dagger}_{\beta,\br} b^{\beta\dagger}_{\alpha,\br+\bdelta}
+\text{H.c.}
 \right]
 \end{split}
\end{equation}
By lattice translation symmetry, we have $e^{i\Phi^{\br,\br+\bdelta}_{\alpha}}=e^{i\Phi^{\br-\bdelta,\br}_{\alpha}}=e^{-i\Phi^{\br,\br-\bdelta}_{\alpha}}$. This implies,
\begin{equation}
\begin{split}
H_{\alpha\beta}&=
J_{1}
\sum_{\br}
\left[
 2b^{\alpha\dagger}_{\beta,\br} b^{\alpha}_{\beta,\br}
 +
 b^{\beta\dagger}_{\alpha,\br-\bdelta} b^{\beta}_{\alpha,\br-\bdelta}
  +
 b^{\beta\dagger}_{\alpha,\br+\bdelta} b^{\beta}_{\alpha,\br+\bdelta}
 \right]
 \\
 &+
 J_{1}
 \sum_{\br}
 \left[
  e^{-i(\Phi^{\br,\br+\delta}_{\beta}-\Phi^{\br,\br+\bdelta}_{\alpha})}
  b^{\alpha\dagger}_{\beta,\br} b^{\beta\dagger}_{\alpha,\br-\bdelta}
  +
    e^{i(\Phi^{\br,\br+\delta}_{\beta}-\Phi^{\br,\br+\bdelta}_{\alpha})}
  b^{\alpha\dagger}_{\beta,\br} b^{\beta\dagger}_{\alpha,\br+\bdelta}
+\text{H.c.}
 \right]
 \end{split}
\end{equation}

Third, we define the Fourier transforms, $b^{\alpha}_{\beta,\br}=(N/4)^{-1/2}\sum_{\bk\in\text{RBZ}}b^{\alpha}_{\beta,\bk}e^{i\bk\cdot\br}$ where the site $\br$ is on
the $\Lambda_{\alpha}$-sublattice. Moreover, $N$ is the number of lattice unit cells and $\bk$ is a momentum in the reduced Brillouin zone of the four-sublattice ordered spin-valley density wave state. We now rewrite the Hamiltonian as,
\begin{align}
H_{\alpha\beta}&=
2J_{1}
\sum_{\bk\in\text{RBZ}}
\left[b^{\beta\dag}_{\alpha,\bk}b^{\beta}_{\alpha,\bk}+b^{\alpha\dag}_{\beta,-\bk}b^{\alpha}_{\beta,-\bk}+
\gamma^{\alpha\beta}_{\bk}b^{\alpha\dag}_{\beta,-\bk}b^{\beta\dag}_{\alpha,\bk}
+
(\gamma^{\alpha\beta}_{\bk})^{*}b^{\alpha}_{\beta,-\bk}b^{\beta}_{\alpha,\bk}\right],
\end{align}
where we have introduced the factor,
\begin{equation}
\gamma^{\alpha\beta}_{\bk}=\cos(\bk\cdot\bdelta
+
\Phi_{\alpha}-\Phi_{\beta}
)
\end{equation}

Fourth, we allow for $J_{2}\neq0$ which amounts to replacements,
\begin{align}
H_{\alpha\beta}&\rightarrow
2(J_{1}+J_{2})
\sum_{\bk\in\text{RBZ}}
\left[b^{\beta\dag}_{\alpha,\bk}b^{\beta}_{\alpha,\bk}+b^{\alpha\dag}_{\beta,-\bk}b^{\alpha}_{\beta,-\bk}+
\gamma^{\alpha\beta}_{\bk}b^{\alpha\dag}_{\beta,-\bk}b^{\beta\dag}_{\alpha,\bk}
+
(\gamma^{\alpha\beta}_{\bk})^{*}b^{\alpha}_{\beta,-\bk}b^{\beta}_{\alpha,\bk}\right],
\\
\gamma^{\alpha\beta}_{\bk}&\rightarrow[J_{1}\cos(\bk\cdot\bR^{(1)}_{\alpha\beta}
+
\Phi_{\alpha}-\Phi_{\beta}
)+
J_{2}\cos(\bk\cdot\bR^{(2)}_{\alpha\beta}
)
]
/(J_{1}+J_{2})
\end{align}
Here, we have also replaced $\bdelta\rightarrow\br^{(1)}_{\alpha\beta}$ and $\br^{(2)}_{\alpha\beta}$ is a vector that connects the second nearest-neighbor sites of the $\Lambda_{\alpha}$- and $\Lambda_{\beta}$-sublattices.

Finally, by performing a Bogoliubov transformation, we arrive at the dispersions
\begin{equation}
\omega^{\alpha\beta}_{\bk}
=
2(J_{1}+J_{2})
\sqrt{1-|\gamma^{\alpha\beta}_{\bk}|^{2}}.
\end{equation}
This concludes the derivation.

\begin{figure}[!b] \centering
\includegraphics[width=0.8\linewidth] {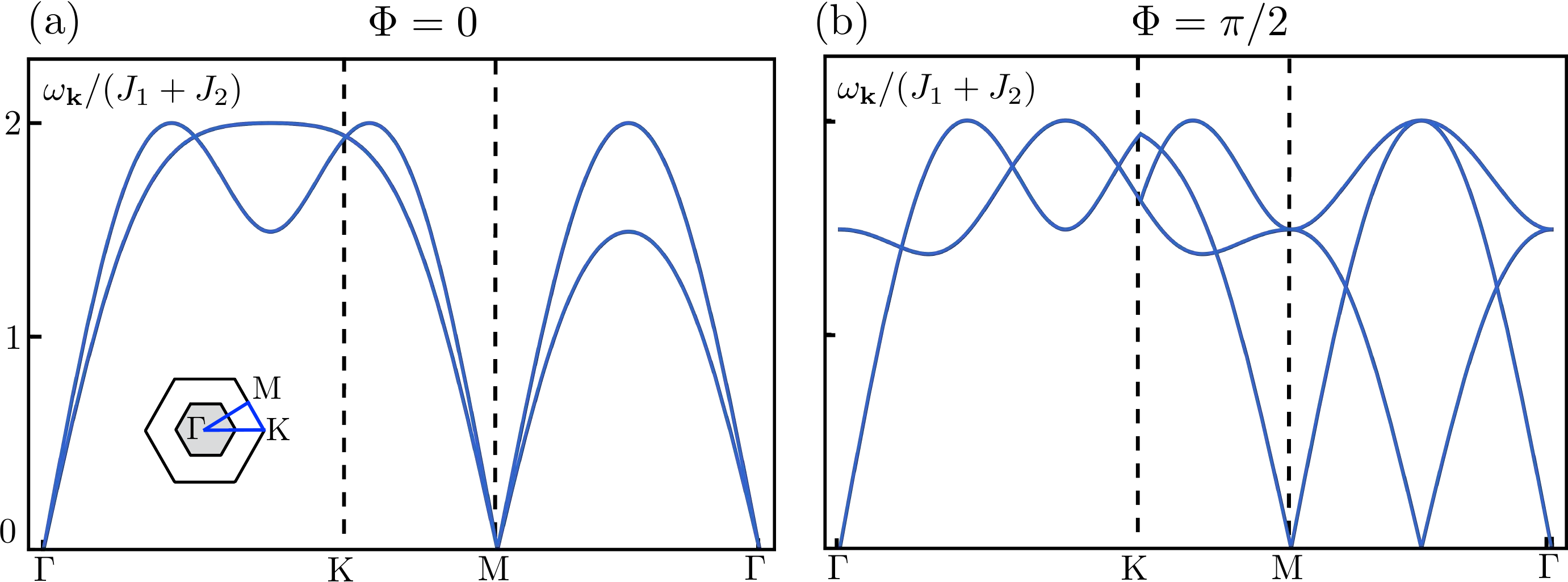}
\caption{(Color online)
(a) Plot of the dispersions relations of Eq.~26 along the high-symmetry lines of the Brillouin zone for $\Phi=0$ and $J_{2}/J_{1}=0.2$. Inset: Brillouin zone of the triangular lattice (white) and structural Brillouin of the four-sublattice spin-valley density wave state (gray). 
(b) Same as (a) but for $\Phi=\pi/2$. 
}\label{fig:sm1}
\end{figure}

\begin{figure}[!b] \centering
\includegraphics[width=0.5\linewidth] {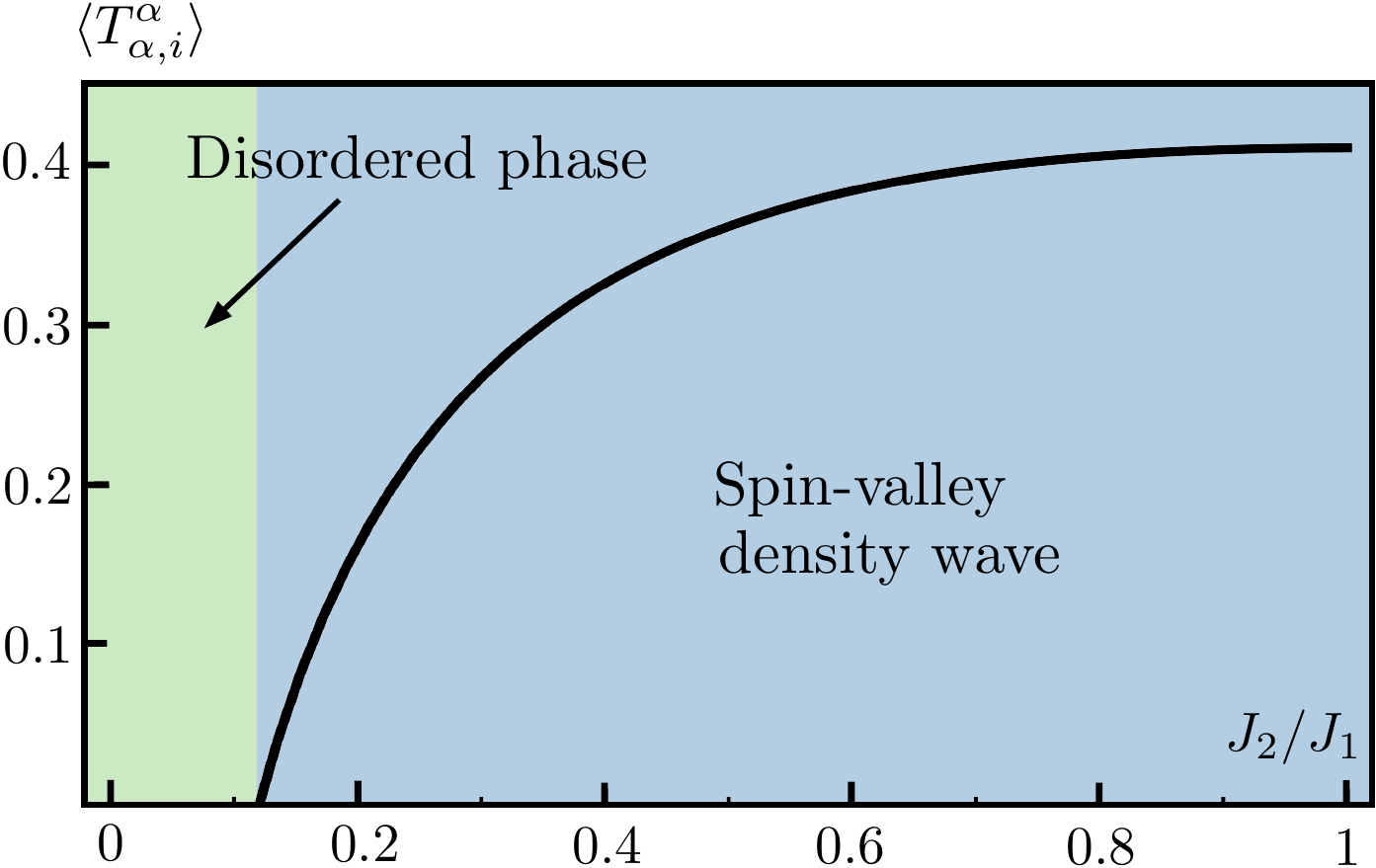}
\caption{(Color online)
Plot of the ordered moment as a function of $J_{2}/J_{1}$. We find that $\langle T^{\alpha}_{\alpha,i}\rangle>0$ for $J_{2}/J_{1}\gtrapprox 0.12$ which implies that the spin-valley density wave order is stabilized in this regime (shown in green). If $J_{2}/J_{1}\lessapprox 0.12$, the spin-valley density wave ordered is destroyed by low-energy quantum fluctuations and the system is in a disordered phase. These results are independent of the value of valley contrasting flux.
}\label{fig:sm2}
\end{figure}

\end{widetext}

\end{document}